\documentclass[sigconf,natbib=false]{acmart}
\usepackage{amsmath,amsfonts}
\usepackage{algorithmic}
\usepackage{graphicx}
\usepackage{textcomp}
\usepackage{xcolor}
\usepackage{color}
\usepackage[dvipsnames]{xcolor}
\usepackage{float}
\usepackage{hyperref}
\usepackage[most]{tcolorbox}
\usepackage{booktabs}
\usepackage{array}
\usepackage{caption}
\usepackage{subcaption}
\usepackage{comment}
\usepackage{ulem}
\usepackage{multirow}

\AtBeginDocument{%
  }

\setcopyright{acmlicensed}

\copyrightyear{2026}
\acmYear{2026}
\acmDOI{XXXXXXX.XXXXXXX}
\acmConference[SAC'26]{The 41st ACM/SIGAPP Symposium on Applied Computing}{March 23--27, 2026}{Thessaloniki, Greece}
\acmISBN{979-X-XXXX-XXXX-X/26/03}

\RequirePackage[
  datamodel=acmdatamodel,
  style=acmnumeric,
  ]{biblatex}

\copyrightyear{2026}
\acmYear{2026}
\setcopyright{cc}
\setcctype{by}
\acmConference[SAC '26]{The 41st ACM/SIGAPP Symposium on Applied Computing}{March 23--27, 2026}{Thessaloniki, Greece}
\acmBooktitle{The 41st ACM/SIGAPP Symposium on Applied Computing (SAC '26), March 23--27, 2026, Thessaloniki, Greece}
\acmPrice{}
\acmDOI{10.1145/3748522.3779726}
\acmISBN{979-8-4007-2294-3/2026/03}

\begin{document}

\title{From Description to Score: Can LLMs Quantify Vulnerabilities?}
\renewcommand{\shorttitle}{}

\author{Sima Jafarikhah}
\affiliation{%
  \institution{University of North Carolina Wilmington}
  \city{Wilmington, NC}
  \country{USA}}
\email{jafarikhaht@uncw.edu}

\author{Daniel Thompson}
\affiliation{%
  \institution{University of North Carolina Wilmington}
  \city{Wilmington, NC}
  \country{USA}}
\email{dbt9576@uncw.edu}

\author{Eva Deans}
\affiliation{%
  \institution{University of North Carolina Wilmington}
  \city{Wilmington, NC}
  \country{USA}}
\email{ecd7121@uncw.edu}

\author{Hossein Siadati}
\affiliation{%
  \institution{University of North Carolina Wilmington}
  \city{Wilmington, NC}
  \country{USA}}
 \email{s.h.siadaty@gmail.com} 

\author{Yi Liu}
\affiliation{%
  \institution{University of North Carolina Wilmington}
  \city{Wilmington, NC}
  \country{USA}}
\email{liuyi@uncw.edu}


\renewcommand{\shortauthors}{Jafarikhah et al.}

\begin{abstract}
Manual vulnerability scoring, such as assigning Common Vulnerability Scoring System (CVSS) scores, is a resource-intensive process that is often influenced by subjective interpretation. This study investigates the potential of general-purpose large language models (LLMs), namely ChatGPT, Llama, Grok, DeepSeek, and Gemini, to automate this process by analyzing over 31{,}000 recent Common Vulnerabilities and Exposures (CVE) entries. The results show that LLMs substantially outperform the baseline on certain metrics (e.g., \textit{Availability Impact}), while offering more modest gains on others (e.g., \textit{Attack Complexity}). Moreover, model performance varies across both LLM families and individual CVSS metrics, with ChatGPT-5 attaining the highest precision. Our analysis reveals that LLMs tend to misclassify many of the same CVEs, and ensemble-based meta-classifiers only marginally improve performance. Further examination shows that CVE descriptions often lack critical context or contain ambiguous phrasing, which contributes to systematic misclassifications. These findings underscore the importance of enhancing vulnerability descriptions and incorporating richer contextual details to support more reliable automated reasoning and alleviate the growing backlog of CVEs awaiting triage.

\end{abstract}

\ccsdesc[500]{Security and privacy~Software and application security}
\ccsdesc[400]{Security and privacy~Vulnerability management}
\ccsdesc[300]{Computing methodologies~Natural language processing}
\ccsdesc[200]{Computing methodologies~Machine learning approaches}

\keywords{Vulnerability, CVSS, CVE, Generative AI}

\maketitle

\section{Introduction} \label{s:introduction}

Vulnerability management is a fundamental component of software security programs across organizations. Public databases such as the National Vulnerability Database (NVD), maintained by NIST’s Information Technology Laboratory (ITL)~\cite{NVD}, catalog and score newly disclosed vulnerabilities. In 2024, the NVD published 40{,}009 CVEs, over 38\% more than in 2023~\cite{gamblin2025cvedata}, increasing the demand for timely and consistent CVSS scoring~\cite{FIRST-CVSS}, which organizations rely on to prioritize remediation and allocate security resources.

This rapid growth has placed substantial strain on maintainers and threat intelligence providers. Throughout 2024, the NVD faced significant processing backlogs~\cite{munshaw2024nvd}, leaving many vulnerabilities unscored or inconsistently characterized and limiting organizations’ ability to assess risk effectively. Internally discovered vulnerabilities pose similar challenges, as they often lack standardized severity ratings, making prioritization subjective and resource-intensive.

On the other hand, recent advances in GenAI have shown that LLMs possess capabilities that go well beyond natural language generation. In particular, they have proven effective in classification tasks, including multi-class scenarios, by reframing classification as a text-to-text task~\cite{gretz2023zero, zhang2025pushing}. This allows LLMs to assign class labels based on textual input without requiring modifications to their underlying architecture \cite{brown2020language, raffel2020exploring}. These advances prompt a natural question: to what extent can generative AI be leveraged to classify vulnerability metrics, such as those defined in the CVE system, using only the textual description of the vulnerability. If effective, such an approach could enable automated vulnerability scoring and help mitigate issues like the backlog observed in the NVD. Motivated by this question, the main contributions of this paper are as follows:
\begin{enumerate}
\item \textbf{Individual LLM Classifiers:} We investigate the feasibility of leveraging general-purpose generative AI models to automate vulnerability scoring by systematically evaluating their ability to assign CVSS metric scores to CVEs using only the textual ``descriptions'' field. It is important to emphasize that including CVE identifiers (CVE ID), would compromise the validity of this evaluation and constitute poor feature engineering, as it allows the LLM to act merely as a retrieval system rather than performing genuine classification.

 
\item \textbf{Meta-LLM Classifiers:} We evaluate multiple LLM models and analyze inconsistencies in their classifications for this task. Furthermore, we construct a meta-classifier to examine whether combining their outputs yields improved performance, or if the observed limitations stem from the inherent ambiguity and contextual insufficiency of existing vulnerability descriptions.
\end{enumerate}

The remainder of this paper is organized as follows. Section~\ref{s:background} reviews the CVSS, the challenges faced by the NVD, and the role of generative AI in classification tasks. Section~\ref{s:rw} reviews and summarizes research on automated CVSS prediction and AI-assisted vulnerability assessment. Section~\ref{s:studyDesign} details the study design, including data collection, model selection, and prompt engineering. Section~\ref{s:evalMetrics} outlines our evaluation metrics, and Section~\ref{s:analysis} presents our findings for both individual LLMs and the proposed meta-classifier. Section~\ref{s:findings} highlights key insights, Section~\ref{s:discuss} discusses implications and future directions, and finally Section~\ref{s:conclude} concludes the study, reflecting on the potential of generative AI to augment vulnerability management workflows.
\section{Background} \label{s:background}
\subsection{Common Vulnerability Scoring System }
The Common Vulnerabilities and Exposures (CVE) program, managed by MITRE, standardizes the identification of publicly known cybersecurity vulnerabilities using unique identifiers. CVE Numbering Authorities (CNAs)—spanning government, industry, and academia—are responsible for assigning and scoring these vulnerabilities to ensure consistent and timely reporting.

The Common Vulnerability Scoring System (CVSS), developed by FIRST.Org, is the leading framework for assessing software vulnerability severity on a scale from 0 (low) to 10 (high). It provides a standardized, open method focused on intrinsic severity, not contextual risk.

\begin{figure}[h]
    \centering
    \includegraphics[scale=.2]{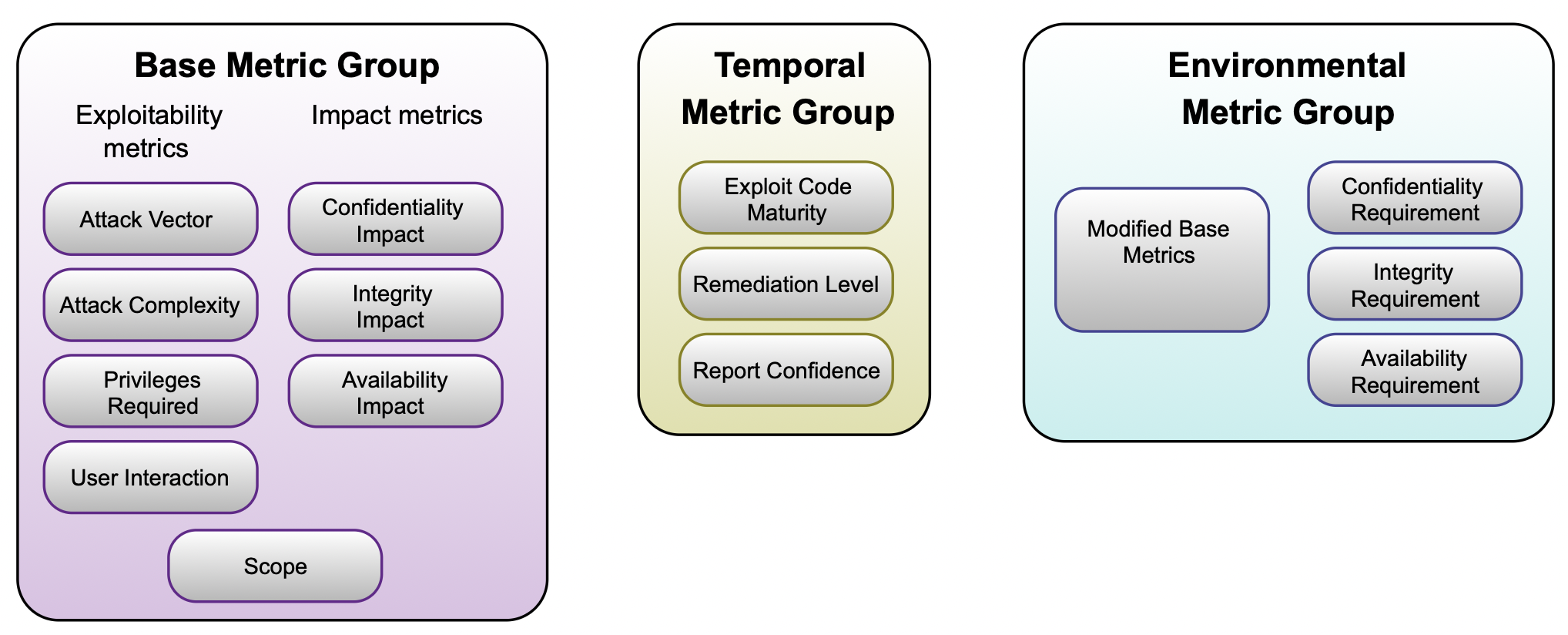}
  \caption{CVSS Metric Groups~\cite{cvssv3.1}}
    \label{fig:metrics_groups}
\end{figure}

As shown in Figure~\ref{fig:metrics_groups}, CVSS is composed of three core metric groups:

\begin{itemize}
  \item \textbf{Base:} Captures the intrinsic properties of a vulnerability. It includes:
  \begin{itemize}
    \item \textbf{Exploitability:} Assesses exploit ease using \textit{Attack Vector (AV)}, \textit{Attack Complexity (AC)}, \textit{Privileges Required (PR)}, and \textit{User Interaction (UI)}. For example, remote vectors (0.85) are more severe than physical ones (0.20).
    \item \textbf{Impact:} Measures effects on \textit{Confidentiality}, \textit{Integrity}, and \textit{Availability (CIA)}—rated as High (0.56), Low (0.22), or None (0.00).
  \end{itemize}
  The Exploitability subscore is:
\begin{equation*}
\text{Exploitability} = 8.22 \times AV \times AC \times PR \times UI
\end{equation*}

  \item \textbf{Temporal:} Adjusts the Base score for factors like exploit maturity or remediation. Multipliers typically range from 0.91 to 1.00.

  \item \textbf{Environmental:} Adapts the score to an organization's context by prioritizing CIA elements (e.g., High = 1.5, Low = 0.5).
\end{itemize}

These components collectively yield the final CVSS score (0–10), allowing organizations to assess severity and prioritize remediation based on both technical and business impact. Table~\ref{tab:cvss_base_metrics_pretty} presents the weights assigned to each class within the individual CVSS metrics.

\begin{table*}[ht]
\centering
\renewcommand{\arraystretch}{1.3}
\begin{tabular}{@{} l c p{7.5cm} l @{}}
\toprule
\textbf{Metric Name} & \textbf{Abbreviation} & \textbf{Possible Values (with numeric scores)} & \textbf{Type} \\
\midrule
Attack Vector & AV & 
N (Network, 0.85), A (Adjacent, 0.62), L (Local, 0.55), P (Physical, 0.20) & Exploitability \\
Attack Complexity & AC & 
L (Low, 0.77), H (High, 0.44) & Exploitability \\
Privileges Required & PR & 
N (None, 0.85), L (Low, 0.62 / 0.68), H (High, 0.27 / 0.50)\newline 
(Values depend on Scope being Unchanged/Changed) & Exploitability \\
User Interaction & UI & 
N (None, 0.85), R (Required, 0.62) & Exploitability \\
Scope & S & 
U (Unchanged), C (Changed) & Impact Modifier \\
Confidentiality & C & 
H (High, 0.56), L (Low, 0.22), N (None, 0.00) & Impact \\
Integrity & I & 
H (High, 0.56), L (Low, 0.22), N (None, 0.00) & Impact \\
Availability & A & 
H (High, 0.56), L (Low, 0.22), N (None, 0.00) & Impact \\
\bottomrule
\end{tabular}
\caption{CVSS v3.1 Base Metrics with Type Classification and Metric Values}
\label{tab:cvss_base_metrics_pretty}
\end{table*}

\subsection{Challenges of Triaging CVEs by NVD}
The NVD has faced challenges in handling the surge of reported vulnerabilities, 28{,}818 CVEs in 2023~\cite{gamblin2025cvedata}, overwhelming its 21-person team~\cite{cybersecuritydive2024cveanalysis}.
By mid-2024, fewer than 10\% of 12,700 CVEs since February had been analyzed \cite{cipollone2024nvdcrisis}, delaying severity scores and product mappings vital to risk assessments \cite{vaughannichols2024nvdslowdown}. Contributing factors include:
\begin{itemize}
    \item \textbf{CVE surge:} Submissions have outpaced manual workflows.
    \item \textbf{Schema migration:} Transition to JSON schema stalled pipelines \cite{nist2024nvdnews}.
    \item \textbf{Submission quality:} Low-quality and duplicate entries consume analyst time.
    \item \textbf{Staffing:} No growth despite rising disclosure volume.
\end{itemize}

These challenges expose systemic weaknesses. While funding and coordination are needed, modernizing and automating NVD processes, including automated vulnerability scoring, is critical to restoring timely vulnerability triage \cite{therecord2024nistbacklog}.

\section{Related Work}\label{s:rw}

\noindent\textbf{Challenges of Vulnerability Scoring.}
Manual vulnerability scoring is prone to inconsistency, as shown by Wunder et al.~\cite{wunder2024shedding}, who found that 68\% of participants changed their CVSS ratings when reassessing the same vulnerabilities. Key metrics like \textit{Attack Vector} and \textit{Privileges Required} were especially error-prone. These results highlight the subjectivity of human scoring and support the need for automated approaches, such as those based on  GenAI, to ensure more consistent and scalable assessments. Stoker et al.~\cite{threatt2024some} showed that despite long-term trends in CVE data, CVSS ratings remained clustered around medium severity, often failing to reflect contextual risk. Similarly, Stoker et al.~\cite{nichols2021heuristic} found inconsistencies in how vendors communicate risk. These findings motivate our use of  GenAI to deliver standardized and consistent scoring.

\noindent\textbf{ML and GenAI-based Scoring of  Vulnerabilities.} Recent studies have explored ML for creating explainable AI to help Counting Number Authority scoring vulnerabilities faster and more accurately. Manai et al.~\cite{manai2024helping} used XGBoost with SHAP to produce interpretable CVSS v3.1 predictions. Yazigi~\cite{yazigi2025modernizing} proposed a GenAI system that aggregates threat data for dynamic risk scoring. Alam et al.~\cite{alam2024ctibench} introduced CTIBench, which benchmarks LLMs on five cybersecurity tasks, including CTI-VSP, closely aligned with our focus. Using zero-shot prompting on 1,000 CVEs, they found performance was sensitive to description length. Our study extends this by analyzing ~31,000+ CVEs, evaluating class imbalance, inter-metric correlations, and model performance using precision, F1-score, and mean absolute error (MAE).
Ullah et al.~\cite{ullah2024secLLMHolmes} conducted a comprehensive evaluation of LLMs for vulnerability detection using their framework SecLLMHolmes. They evaluated eight LLMs, showing that even advanced models exhibit high false positive rates, inconsistent outputs, and fragile reasoning that breaks under minor code augmentations. Their findings highlight that LLMs are not yet robust enough for automated vulnerability detection in real-world settings, underscoring the need for further research into model reliability and reasoning faithfulness. It should be emphasized that the incorporation of CVE identifiers in prompts to LLMs, together with the corresponding description as in prior work~\cite{marchiori2025can}, allows models to retrieve known vectors and undermines true classification.


\section{Study Design}\label{s:studyDesign}

\subsection{Objective} The objective of this study is to systematically evaluate the predictive capabilities of LLMs for automated vulnerability assessment, with a focus on accurately inferring CVSS scores from textual vulnerability descriptions. 

\subsection{Data Collection}

The study utilizes the publicly available CVE List maintained by MITRE~\cite{cve2025list} as its primary data source. While all CVEs from 2004 to 2024 were initially retrieved, the analysis focuses on entries published from 2019 onward to ensure consistency with the CVSS v3.1 standard. After filtering for CVEs with complete CVSS v3.1 base metrics and valid English descriptions, the final dataset consists of 31,000+ entries. Each CVE, stored in JSON format, includes a human-readable description and eight CVSS v3.1 base metrics. Additional quality filters excluded CVEs with missing, non-English, or incomplete data.
For the purposes of this study, the v3.1 CVSS base metrics and descriptions obtained from the CVE List are treated as ground truth. Although there may be some variance in scores and descriptions, the MITRE-maintained CVE List is a widely adopted and standardized source of vulnerability information~\cite{cve_process}.

\subsection{Large Language Models}
This study employs six models: GPT-4o (G4) \cite{openai2024gpt4o}, GPT-5 (G5) \cite{openai2025gpt5}, Llama-3.3-70B-Instruct (L) \cite{meta2024llama3}, Gemini-2.5-Flash (GM) \cite{deepmind2025geminiflash}, DeepSeek-R1 (DS) \cite{deepseek2025r1}, and Grok-3 (GR) \cite{xai2025grok3} to predict CVSS base metrics directly from CVE descriptions. We intentionally selected a mix of high-performing closed-source and open-weight models to capture complementary strengths: closed models such as GPTs typically offer state-of-the-art accuracy, robustness, and safety features, while open models such as Llama provide transparency, reproducibility, and adaptability for security research. These characteristics make them strong candidates for CVSS scoring, which requires both precision and interpretability. We used Azure's AI Foundry for GPT-4o, GPT-5, Llama, DeepSeek, and Grok, and Google's AI Studio for Gemini.

\subsection{Prompt Engineering}  We followed OpenAI’s best practices~\cite{openai2025prompt} in designing the prompt that directed the LLMs to generate CVSS base metric scores. One best practice is to be specific on the task and avoid combining results~\cite{mirzadeh2024gsm}. Therefore, our prompt did not ask for the overall CVSS score and instead asked for itemized scoring of metrics. We instructed the model to act as a cybersecurity expert. The prompt followed a two-step structure: (1) extract the eight base metrics, and (2) output them in a fixed format for consistency. To ensure deterministic outputs, if configurable, the temperature was set to 0, following CTIBench guidance~\cite{alam2024ctibench}. We evaluated zero-, two-, five-, and ten-shot prompts on a subset of 6,000 CVE entries from 2024. The two-shot variant achieved the best performance; therefore, it was adopted for all subsequent experiments. A Python script automated API calls in batches of 20 CVEs, appending predicted metrics to CSV outputs. Using this pipeline, we processed over 31,000 CVE records for downstream analysis.

\section{Evaluation Metrics}\label{s:evalMetrics}
To assess the quality of CVSS scores produced by GPT-4o, GPT-5, Llama-3.3-70B-Instruct, Gemini-2.5-Flash, DeepSeek-R1, and Grok-3, we employ several complementary evaluation metrics. An examination of the CVE dataset reveals a class imbalance, making it important to report class-sensitive metrics such as per-class precision, recall, and F1-score, as well as their macro or weighted averages. These metrics offer more informative insights into the models’ performance across all classes.

\begin{enumerate}
\item \textbf{Accuracy.}  Accuracy measures the proportion of predictions that exactly match the true labels, indicating overall model performance. However, due to class imbalance, it can be misleading and should be interpreted with caution.

\begin{equation*}
\text{Accuracy} = \frac{\# \text{ correct } (y_i = \hat{y}_i)}{N}
\end{equation*}
where  \( y_i \) denotes the true label, and \( \hat{y}_i \) represents the predicted label.

\item\textbf{Precision.} Precision measures the proportion of correct positive predictions out of all predicted positives, reflecting the model’s accuracy in assigning labels. For a specific class \( i \), it is defined as:

\[
\text{Precision}_i = \frac{TP_i}{TP_i + FP_i}
\]

where \( TP_i \) and \( FP_i \) denote the true and false positives for class \( i \), respectively.

To address class imbalance, we compute \textit{weighted precision}, which averages per-class precision values weighted by the number of true instances \( n_i \) in each class:

\[
\text{Weighted Precision} = \frac{1}{N} \sum_{i=1}^{K} n_i \cdot \text{Precision}_i
\]

Here, \( K \) is the total number of classes, and \( N = \sum_{i=1}^{K} n_i \) is the total number of samples. Reporting both per-class and weighted precision ensures a more balanced assessment of model performance across all classes.
\item \textbf{Recall.} Recall is a measure of  ability to identify positive instances. Weighted recall provides an aggregate measure of recall across all classes by accounting for class imbalance. It is computed as the weighted average of per-class recall values, where each class is weighted by its number of true instances (support). For a classification task with \( K \) classes, the weighted recall is defined as:

\[
\text{Weighted Recall} = \frac{1}{N} \sum_{i=1}^{K} n_i \cdot \text{Recall}_i
\]

where:
\begin{itemize}
    \item \( \text{Recall}_i = \frac{TP_i}{TP_i + FN_i} \) is the recall for class \( i \),
    \item \( TP_i \) and \( FN_i \) are the number of true positives and false negatives for class \( i \), respectively,
    \item \( n_i \) is the number of true instances in class \( i \), and
    \item \( N = \sum_{i=1}^{K} n_i \) is the total number of instances across all classes.
\end{itemize}
\item\textbf{F1-Score.} The F1-score is the harmonic mean of precision and recall, capturing a balance between false positives and false negatives. For class \( i \), it is defined as:

\[
\text{F1}_i = 2 \times \frac{\text{Precision}_i \times \text{Recall}_i}{\text{Precision}_i + \text{Recall}_i}
\]

To address class imbalance, we compute the \textit{weighted F1-score}, which averages per-class F1-scores weighted by the number of true instances \( n_i \):

\[
\text{Weighted F1} = \frac{1}{N} \sum_{i=1}^{K} n_i \cdot \text{F1}_i
\]

\item\textbf{Mean Absolute Error (MAE).} Some CVSS metrics, like \textit{AV}, have an inherent ordinal order (e.g., \textit{Network} to \textit{Physical}). To capture the severity of prediction errors, we assign ordinal values and compute MAE as:

\[
\text{MAE} = \frac{1}{N} \sum_{i=1}^{n} | y_i - \hat{y}_i |
\]

where \( y_i \) and \( \hat{y}_i \) are the true and predicted ordinal labels, respectively, and \( N \) is the total number of samples. This metric reflects not just correctness, but also how far predictions deviate from true values.
\end{enumerate}

\section{Analysis}\label{s:analysis}

\subsection{Exploring the CVE Dataset}

\begin{figure}
    \centering

    \begin{subfigure}[b]{0.45\linewidth}
        \includegraphics[width=\linewidth]{Images/cvss-metrics.png}
        \caption{Attack Complexity}
    \end{subfigure}
    \begin{subfigure}[b]{0.45\linewidth}
        \includegraphics[width=\linewidth]{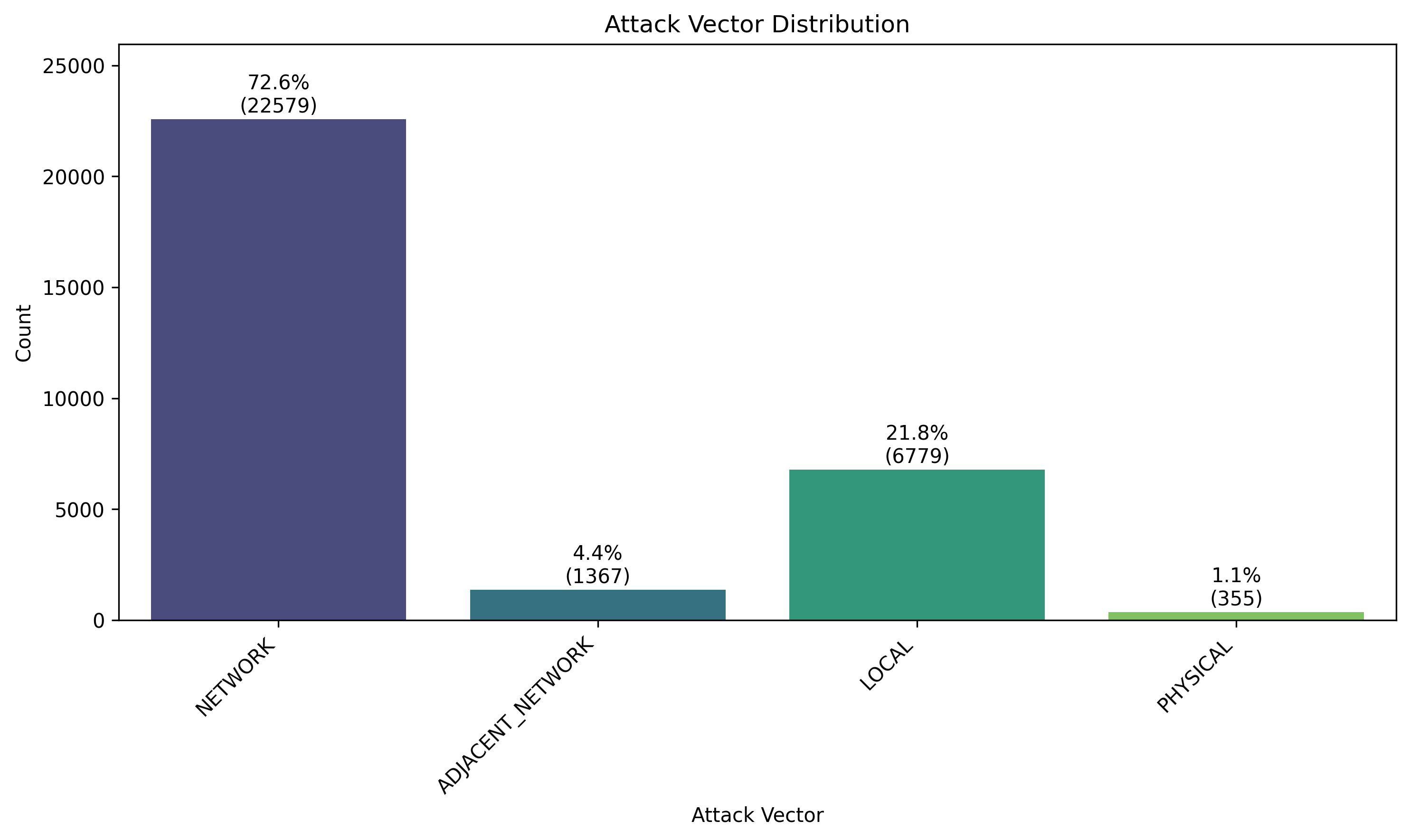}
        \caption{Attack Vector}
    \end{subfigure}

    \begin{subfigure}[b]{0.45\linewidth}
        \includegraphics[width=\linewidth]{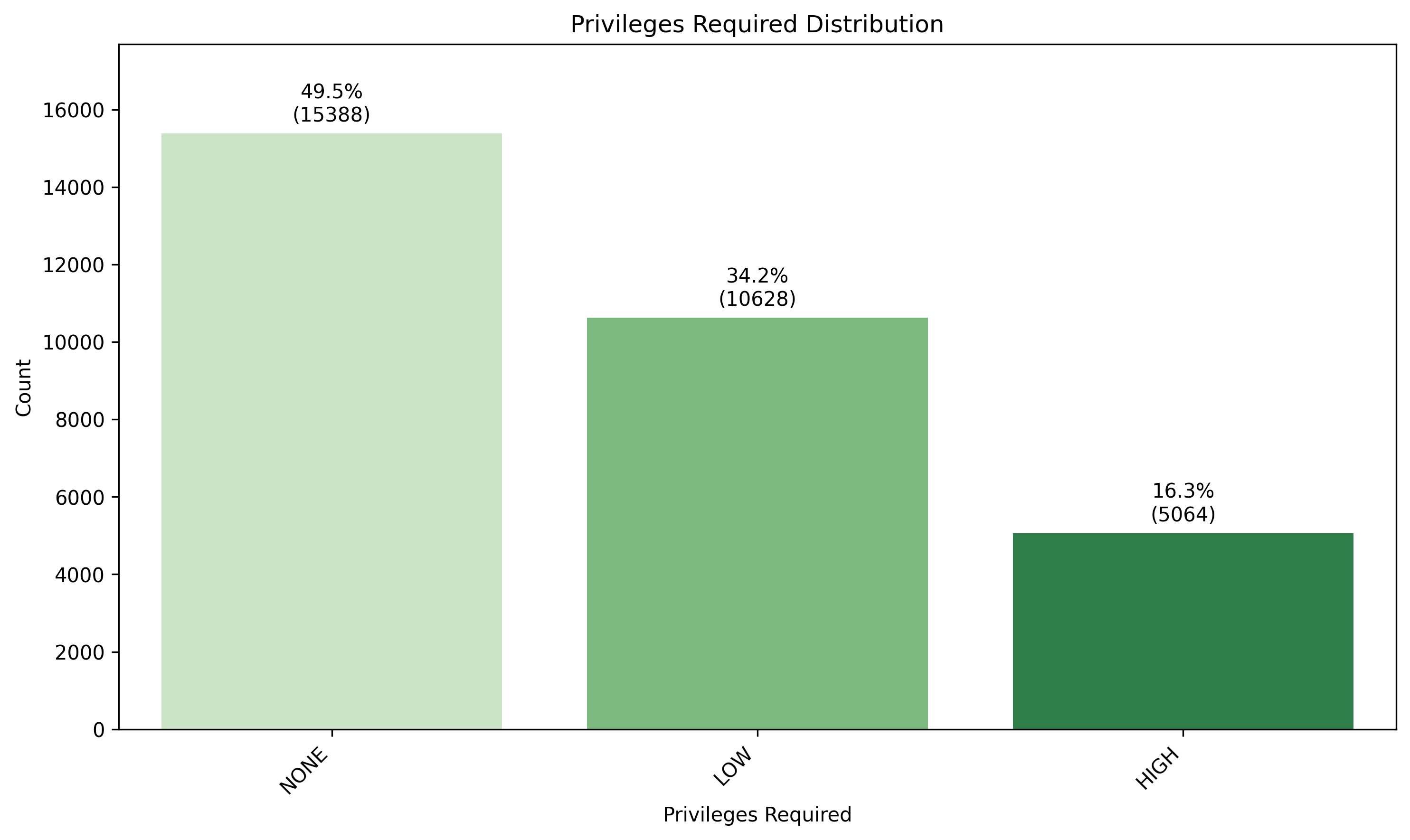}
        \caption{Privileges Required}
    \end{subfigure}
    \begin{subfigure}[b]{0.45\linewidth}
        \includegraphics[width=\linewidth]{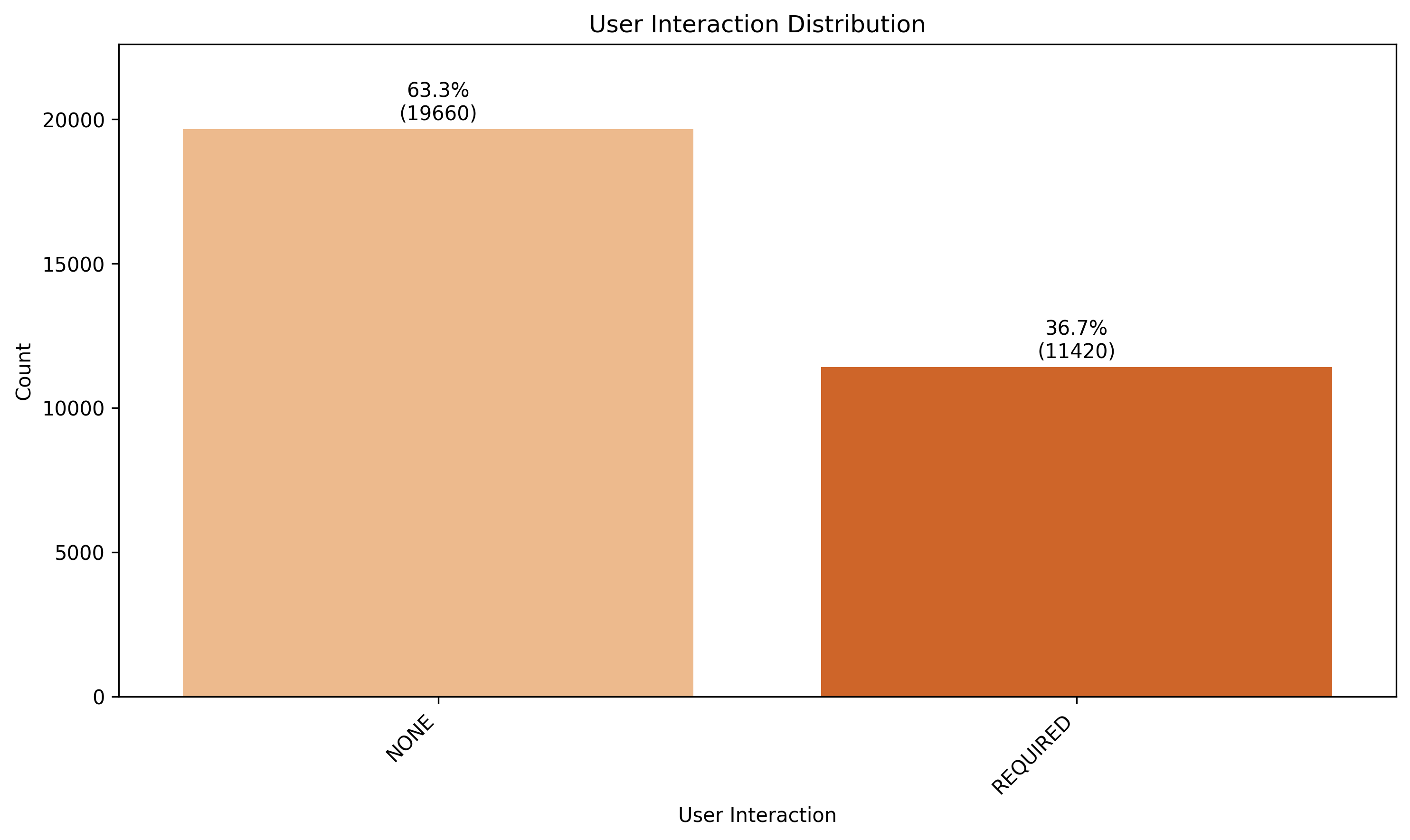}
        \caption{User Interaction}
    \end{subfigure}

    \begin{subfigure}[b]{0.45\linewidth}
        \includegraphics[width=\linewidth]{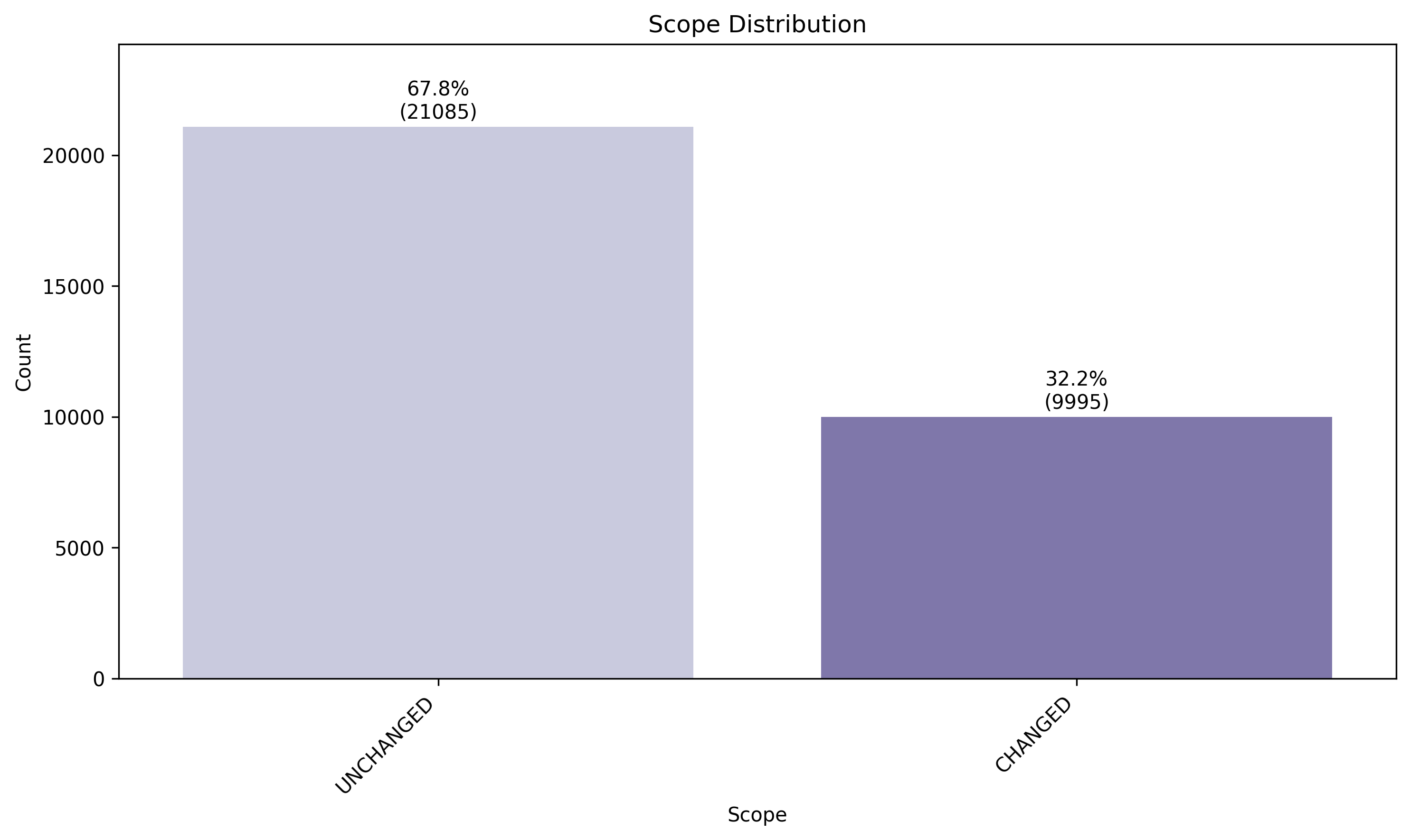}
        \caption{Scope}
    \end{subfigure}
    \begin{subfigure}[b]{0.45\linewidth}
        \includegraphics[width=\linewidth]{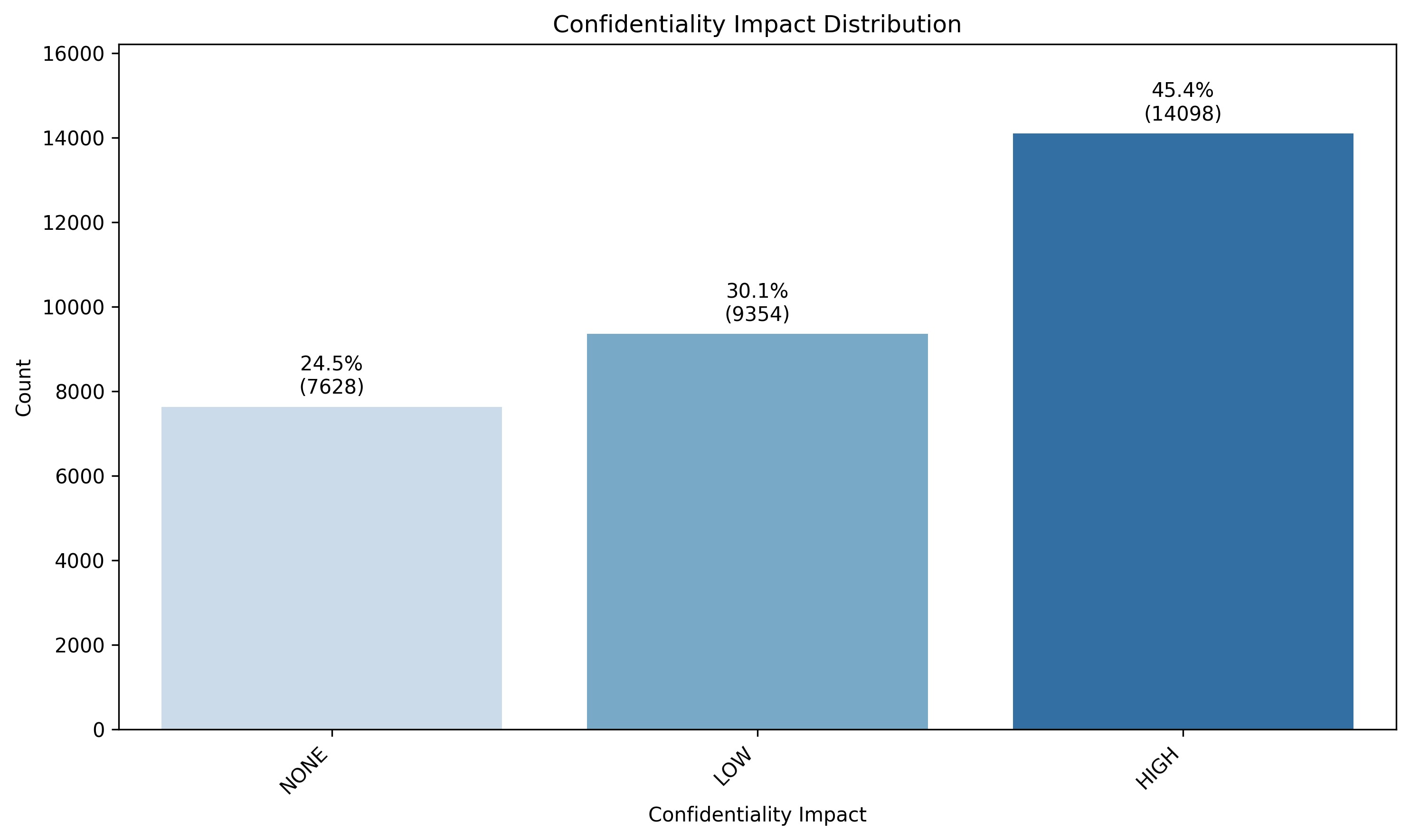}
        \caption{Confidentiality Impact}
    \end{subfigure}

    \begin{subfigure}[b]{0.45\linewidth}
        \includegraphics[width=\linewidth]{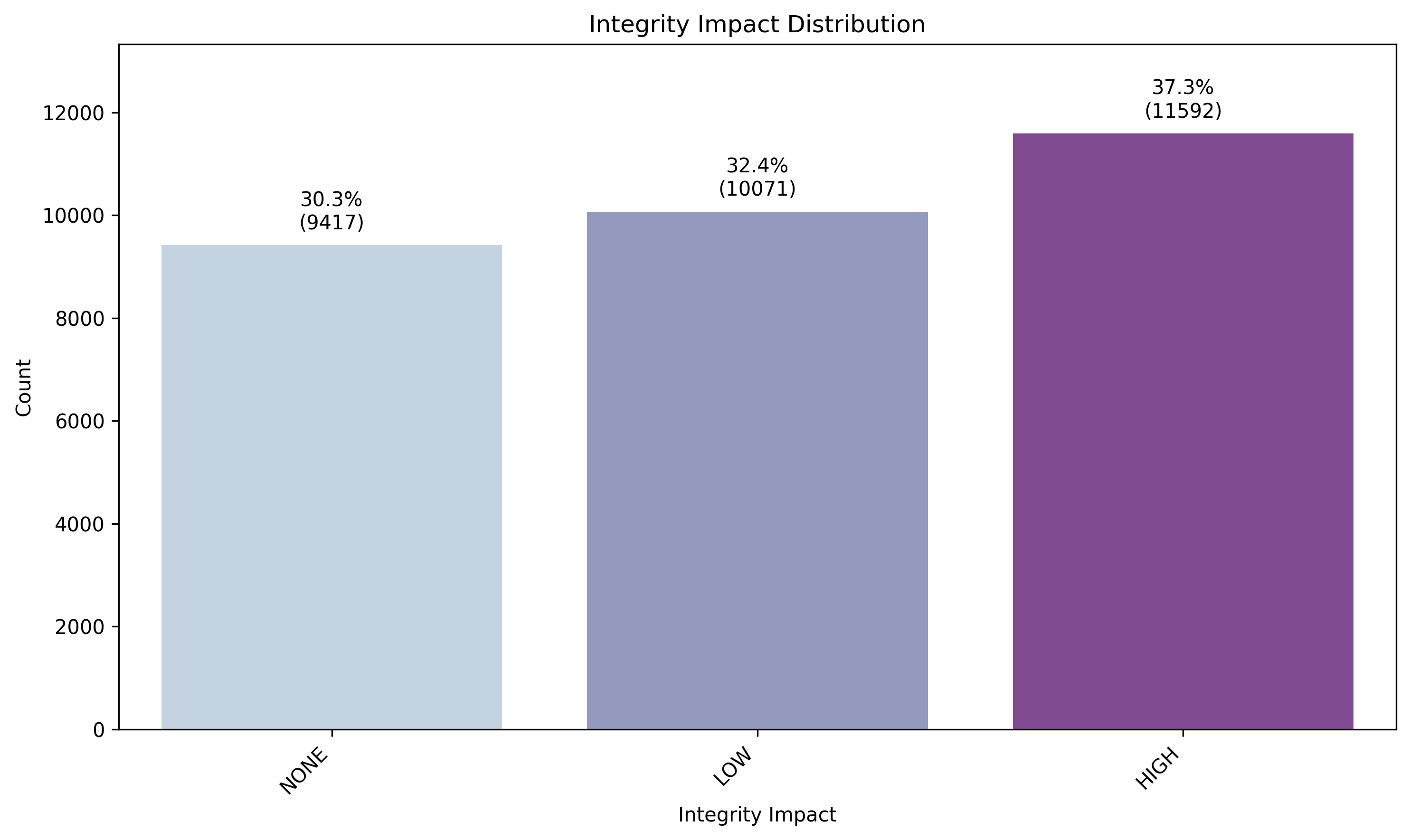}
        \caption{Integrity Impact}
    \end{subfigure}
    \begin{subfigure}[b]{0.45\linewidth}
        \includegraphics[width=\linewidth]{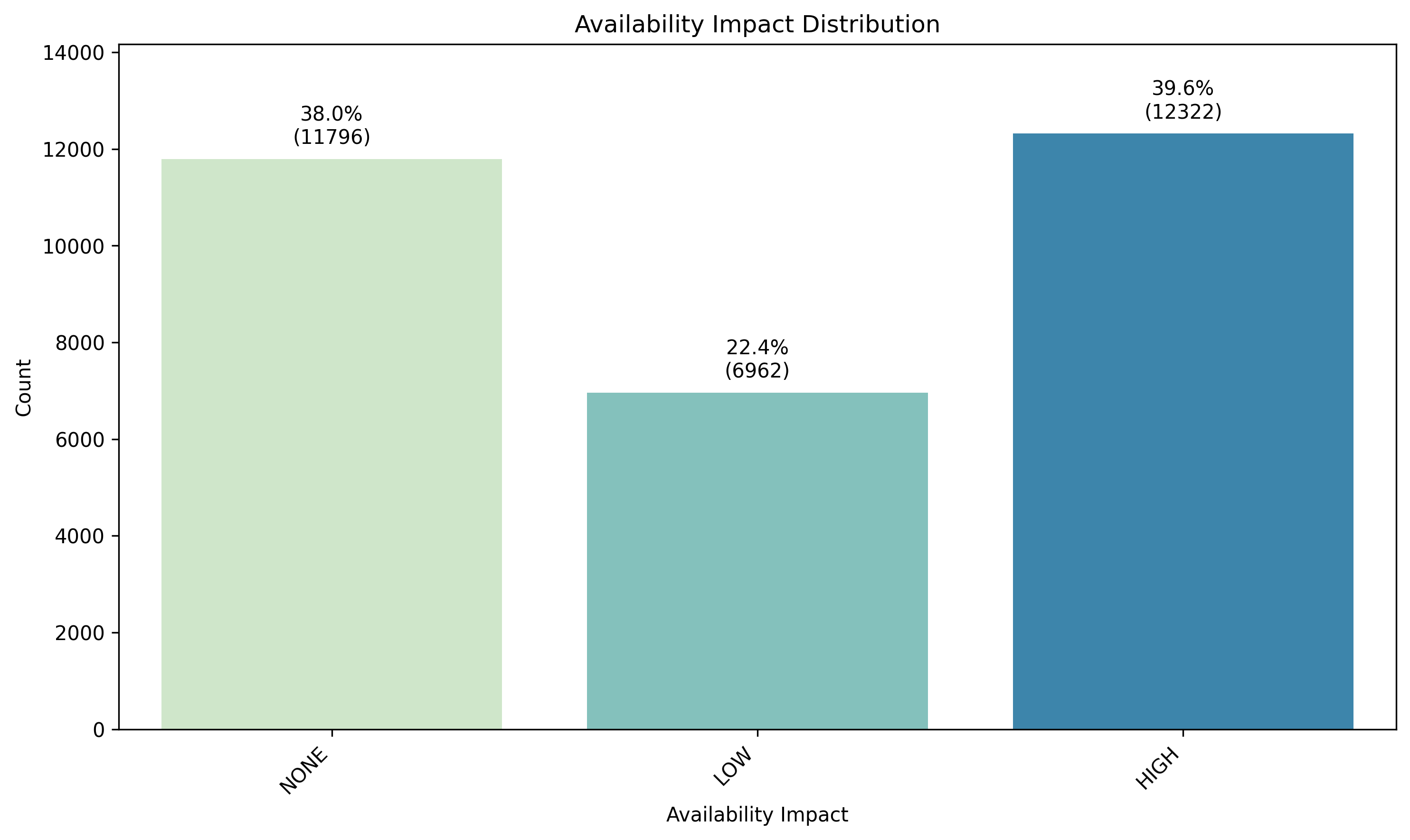}
        \caption{Availability Impact}
    \end{subfigure}

    \caption{Distribution of Categories Across CVSS Metrics}
    \label{fig:cvss_distributions}
\end{figure}

\noindent\textbf{Baseline.}  Figure~\ref{fig:cvss_distributions} illustrates class imbalances across CVSS base metrics. For instance, 83.8\% of vulnerabilities are labeled as \textsc{Low} complexity. To quantify this imbalance, the study uses the imbalance ratio, defined as the number of samples in the majority class divided by those in the minority class. A higher ratio signifies greater disparity, which can negatively affect evaluation metrics like accuracy.


As shown in Table~\ref{tab:security-metrics}, the dataset exhibits notable class imbalance across several metrics. \textit{Attack Vector} is the most imbalanced, with the majority class being 63.60 times larger than the minority. In contrast, \textit{Integrity Impact} is the most balanced among the metrics considered.

\begin{table}
\centering
\begin{tabular}{lr}
\hline
\textbf{Metric} & \textbf{Imbalance Ratio} \\
\hline
Attack Complexity         & 5.19 \\
Attack Vector             & 63.60 \\
Privileges Required       & 3.04 \\
User Interaction          & 1.72 \\
Scope                     & 2.11 \\
Confidentiality Impact    & 1.85 \\
Integrity Impact          & 1.23 \\
Availability Impact       & 1.77 \\
\hline
\end{tabular}
\caption{Imbalance Ratio of CVSS metrics}
\label{tab:security-metrics}
\end{table}

\begin{figure}[H]
    \centering
    \includegraphics[width=\linewidth]{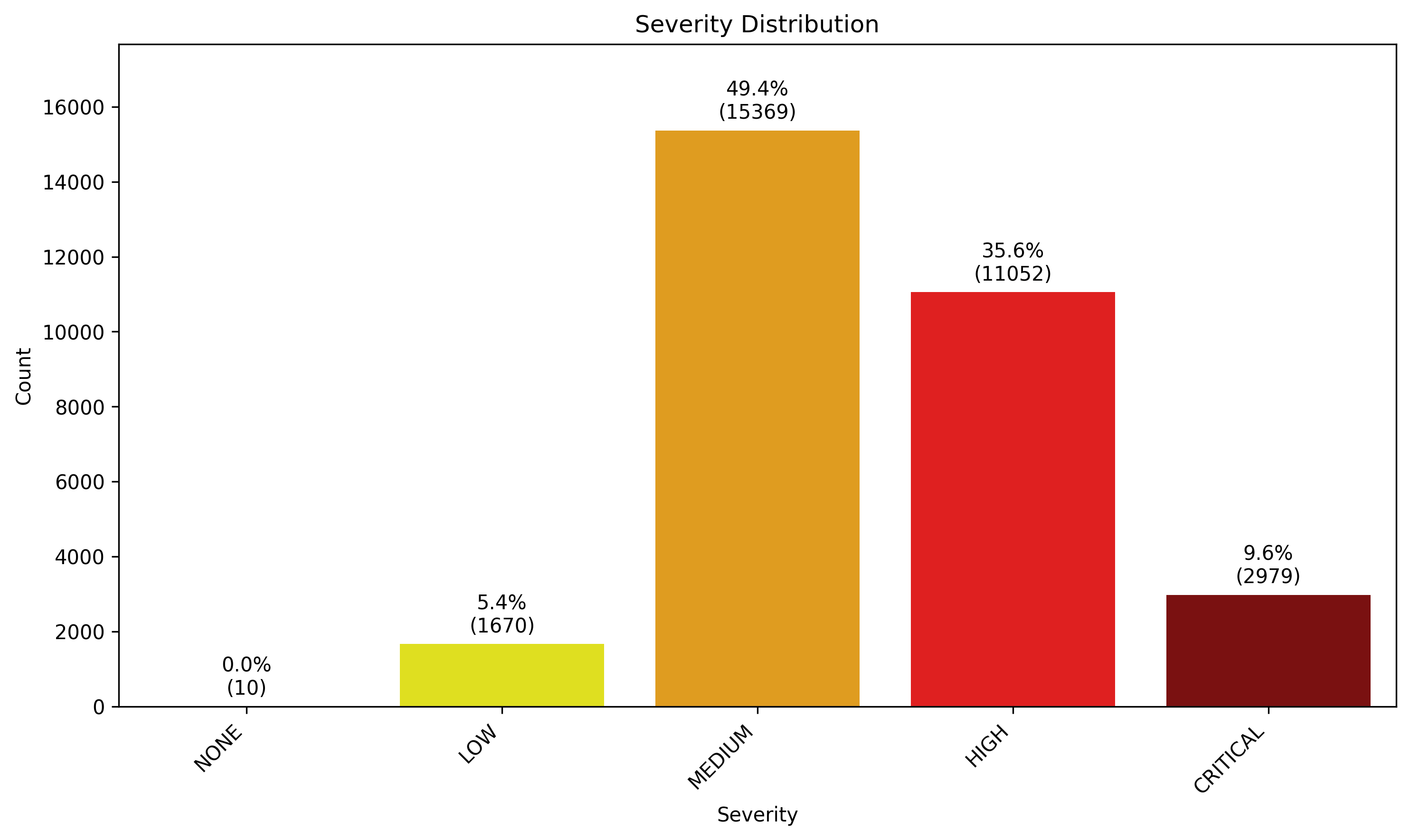}
    \caption{Distribution of Categories for Severity}
    \label{fig:sd}
\end{figure}

\noindent\textbf{Distribution of severity classes.} 
Figure~\ref{fig:sd} illustrates the distribution of severity scores. The presence of class imbalance is evident and should be considered during the analysis.

\begin{figure}[H]
    \centering
    \includegraphics[width=\linewidth]{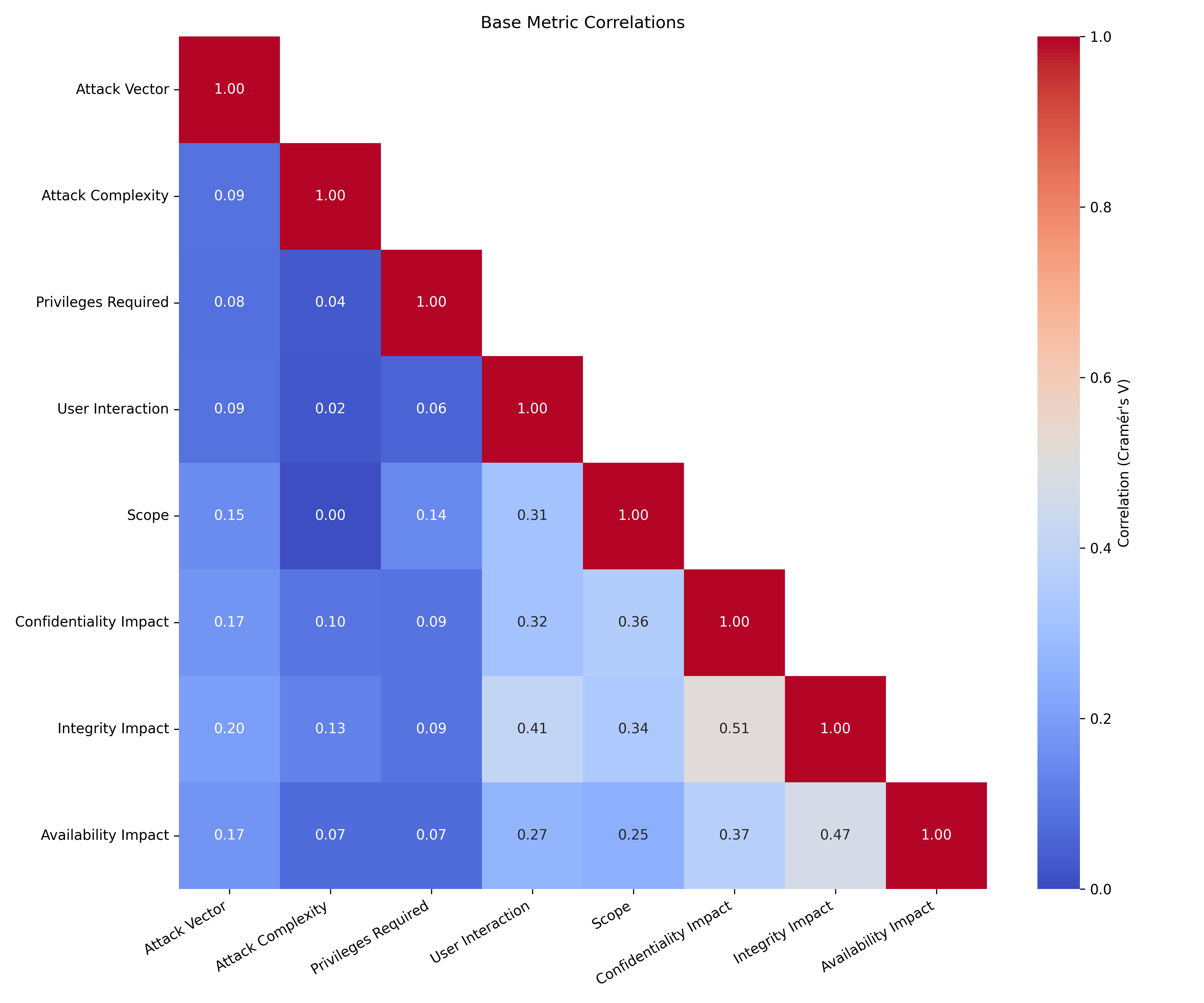}
    \caption{Correlation of Base Metrics}
    \label{fig:metrics_correlation}
\end{figure}

\noindent\textbf{Correlation analysis.}
We use Cramer's V to assess relationships between categorical CVSS metrics. Results indicate a moderate to strong correlation between Integrity and Confidentiality, with moderate correlations also observed between User Interaction and Integrity, and between Integrity and Availability. Other metric pairs show weak or negligible associations (See Figure ~\ref{fig:metrics_correlation}).

\begin{figure*}[htbp]
    \centering

    \begin{subfigure}[b]{0.24\textwidth}
        \includegraphics[width=\linewidth]{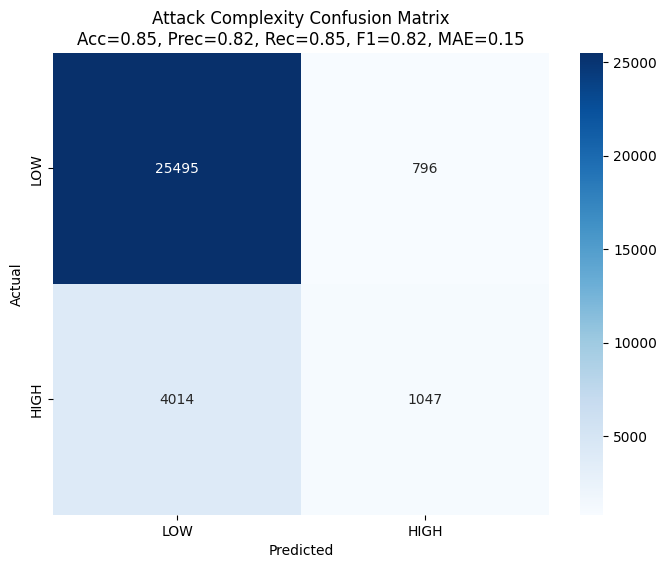}
        \caption{Attack Complexity}
    \end{subfigure}
    \hfill
    \begin{subfigure}[b]{0.24\textwidth}
        \includegraphics[width=\linewidth]{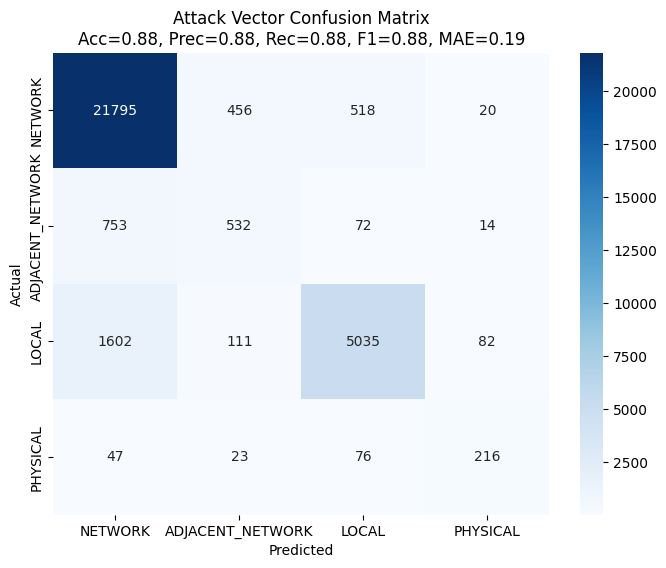}
        \caption{Attack Vector}
    \end{subfigure}
    \hfill
    \begin{subfigure}[b]{0.24\textwidth}
        \includegraphics[width=\linewidth]{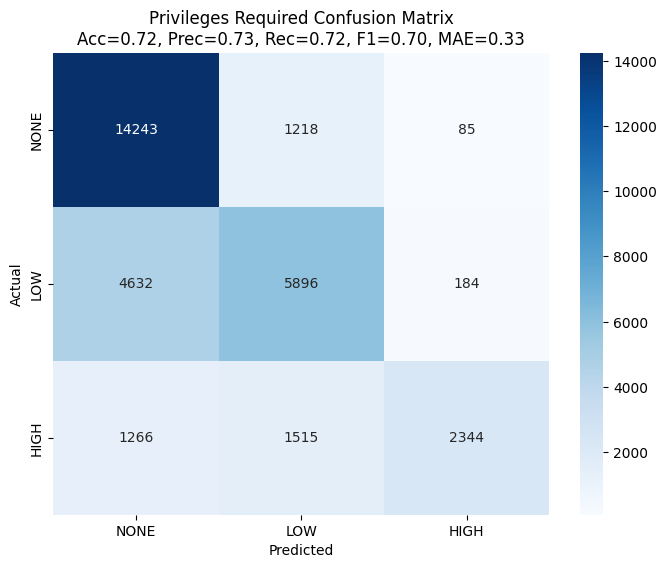}
        \caption{Privileges Required}
    \end{subfigure}
    \hfill
    \begin{subfigure}[b]{0.24\textwidth}
        \includegraphics[width=\linewidth]{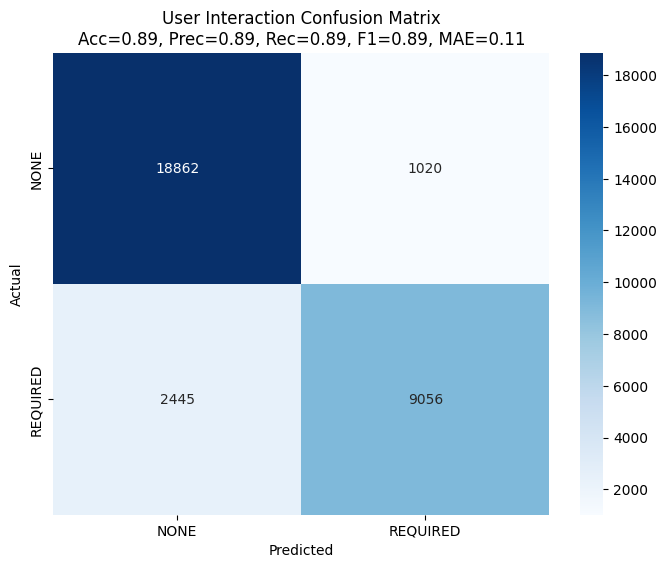}
        \caption{User Interaction}
    \end{subfigure}

    \vspace{1em}

    \begin{subfigure}[b]{0.24\textwidth}
        \includegraphics[width=\linewidth]{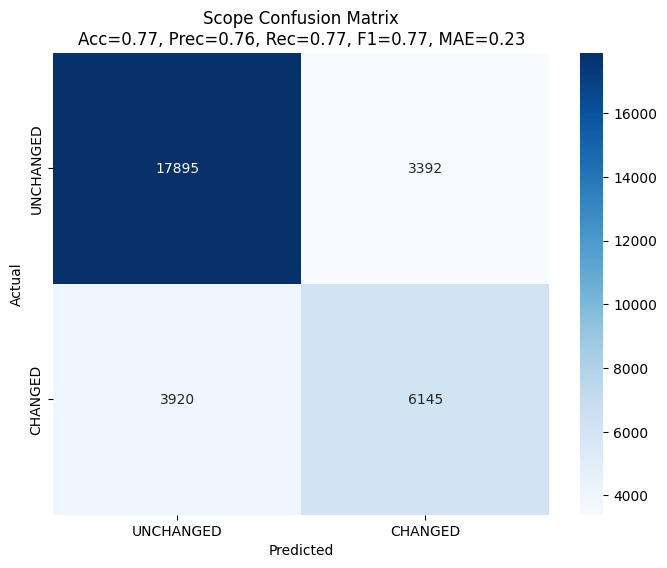}
        \caption{Scope}
    \end{subfigure}
    \hfill
    \begin{subfigure}[b]{0.24\textwidth}
        \includegraphics[width=\linewidth]{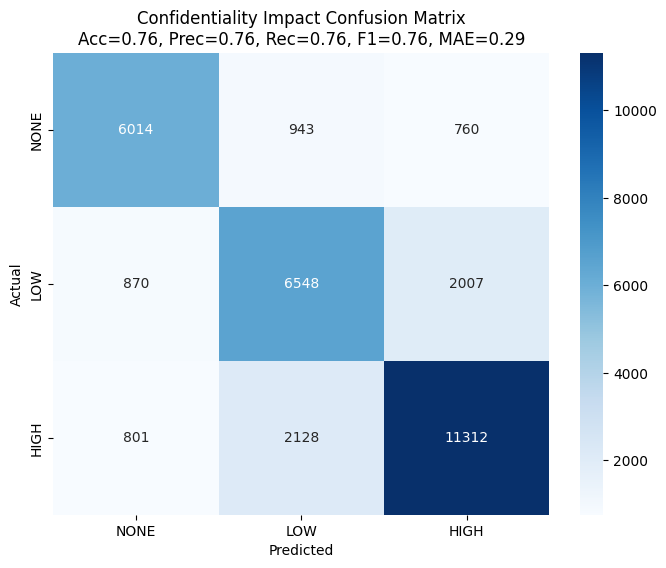}
        \caption{Confidentiality Impact}
    \end{subfigure}
    \hfill
    \begin{subfigure}[b]{0.24\textwidth}
        \includegraphics[width=\linewidth]{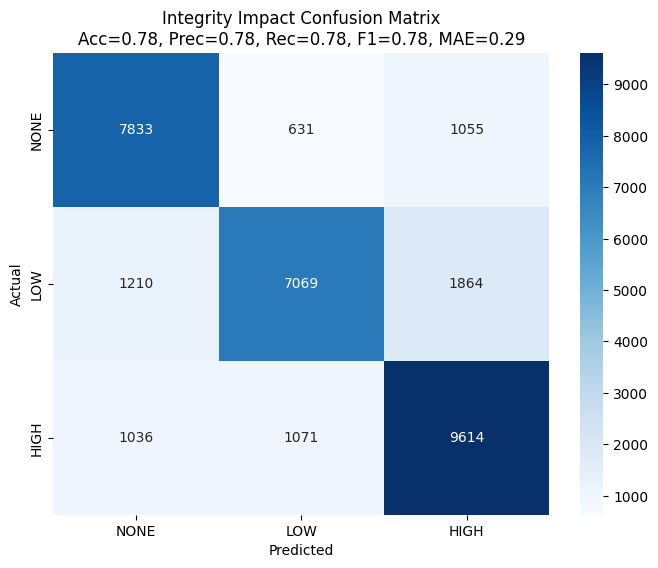}
        \caption{Integrity Impact}
    \end{subfigure}
    \hfill
    \begin{subfigure}[b]{0.24\textwidth}
        \includegraphics[width=\linewidth]{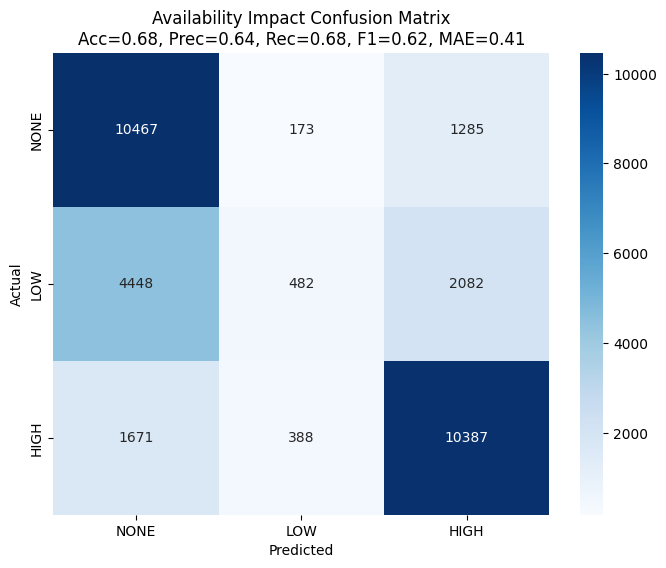}
        \caption{Availability Impact}
    \end{subfigure}

    \caption{Confusion Matrices for CVSS metrics for GPT-5, the best performing model, per metric}
    \label{fig:all_cm_roc}
\end{figure*}

Table~\ref{tab:performance_metrics} summarizes the performance of the LLMs in predicting CVSS metric labels for CVEs. We analyze each metric individually before presenting combined results, using both the performance metrics in Table~\ref{tab:performance_metrics} and the confusion matrices in Figure~\ref{fig:all_cm_roc}.

\subsection{Analysis of Individual LLM Classifiers}
Table~\ref{tab:performance_metrics} presents the detailed performance of individual LLMs across all CVSS and evaluation metrics.\\
\noindent\textbf{Attack Complexity.} GPT-5 achieved the highest accuracy at 84.66\%, slightly above the baseline of 83.85\%, with balanced precision (0.72), strong recall (0.85) and yielding the best F1-score (0.82) between all models. GPT-4o followed closely (81.92\%, F1=0.78) behind. Gemini (80.21\%), DeepSeek (79.29\%), and Grok(80.13\%) all performed comparably (F1=0.78), with Llama's performance trailing behind (77.74\%, F1=0.77).

\noindent\textbf{Attack Vector.} Gemini led with 89.42\% accuracy, surpassing the 72.65\% baseline and achieving balanced scores (P=0.82, R=0.83, F1=0.82). GPT-5 followed at 87.96\% (F1=0.88), showing the strongest recall (0.88) but slightly lower precision. GPT-4o reached 86.57\%, also well above baseline with consistent metrics (F1 = 0.86). Llama (79.31\%) and DeepSeek (79.01\%) showed moderate results $(F1=0.76–0.79)$, while Grok (83.00\%) matched Gemini’s stability (F1 = 0.82). Although all models outperformed the baseline, multiple showed a bias toward the majority class, often predicting NETWORK over minority classes like LOCAL.

\noindent\textbf{Privileges Required.} GPT-5 performed the best, with 71.61\% accuracy in comparison to the 49.51\% baseline, and maintained consistency in other metrics (P=0.76, R=0.72, F1=0.70). Gemini followed (69.74\%, F1=0.63) with GPT-4o (65.44\%, F1=0.64) and Grok (64.75\%, F1=0.63) behind. Llama (58.69\%) and DeepSeek (58.49\%) had the weakest results. There was a pattern of confusion between LOW and NONE levels, as models struggled to differentiate between the two classes.

\noindent\textbf{User Interaction.} GPT-5 achieved the top accuracy (88.95\%) and highest overall balance (P=0.89, R=0.89, F1=0.89), outperforming the 63.26\% baseline. Gemini (87.85\%, F1=0.82) and GPT-4o remained competitive (84.13\%, F1 = 0.84), with Grok (82.58\%, F1=0.82) following behind. DeepSeek (75.18\%, F1=0.73) and Llama (77.67\%, F1=0.77) once again had the weakest performance. Errors frequently involved mislabeling REQUIRED as NONE, showing class bias across models.

\noindent\textbf{Scope.} GPT-5 led with 76.68\% accuracy, above the 67.84\% baseline, achieving F1 = 0.77 with good recall (0.77). Grok (76.06\%) and Gemini (71.25\%) also showed balanced outputs (F1=0.75). GPT-4o performed stably (72.86\%, F1=0.73), while DeepSeek (70.36\%, F1=0.68) and Llama (69.48\%, F1=0.70) trailed. Most models favored the majority UNCHANGED class.

\noindent\textbf{Confidentiality Impact.} GPT-5 attained the best performance (76.05\%) over the 45.36\% baseline, with all metrics at 0.76, indicating consistent predictions. Gemini (74.01\%) also performed accurately, though its other metrics were lower than GPT-5 at 0.65. and GPT-4o (66.96\%) followed, while Grok (65.40\%) and DeepSeek (60.27\%) remained moderate. Llama performed weakest (56.47\%, F1=0.57). There was overall difficulty distinguishing between LOW and HIGH impact, although GPT-5's higher accuracy suggests better contextual understanding.

\noindent\textbf{Integrity Impact.} GPT-5 again outperformed other models with 78.10\% accuracy (F1=0.78), well above the 37.30\% baseline. Gemini (75.38\%) and Grok (68.24\%) followed closely (F1=0.68), while GPT-4o (69.54\%) performed steadily (F1 = 0.69). DeepSeek (56.58\%) and Llama (55.65\%) were the weakest performers.

\noindent\textbf{Availability Impact.} GPT-5 led with 67.95\% accuracy (F1=0.62) versus the 39.65\% baseline, though this was its weakest metric (MAE=0.41). Gemini (66.06\%, F1=0.55) followed. GPT-4o (61.84\%, F1=0.57) and Grok (60.41\%, F1=0.55) showed comparable trends, while DeepSeek (51.86\%) and Llama (51.13\%) underperformed, both exhibiting an MAE >= 0.7. 

\begin{table*}[p]
\centering
\tiny
\renewcommand{\arraystretch}{0.3}
\setlength{\tabcolsep}{3pt}
\resizebox{\textwidth}{!}{%
\begin{tabular}{@{}lccccccc@{}}
\toprule
\textbf{Metric Name} & \textbf{Model} & 
\textbf{Accuracy (\%)} & \textbf{Precision} & \textbf{Recall} & \textbf{F1} & \textbf{MAE} & \textbf{Baseline (\%)} \\
\midrule
\multirow{6}{*}{Attack Complexity}
 & G4 & 81.92 & \textbf{0.77} & 0.82 & 0.78 & 0.18 & 83.85 \\
 & G5 & \textbf{84.66} & 0.72 & \textbf{0.85} & \textbf{0.82} & \textbf{0.15} &  \\
 & L  & 77.74 & 0.75 & 0.79 & 0.77 & 0.44 &  \\
 & GM & 80.21 & \textbf{0.77} & 0.80 & 0.78 & 0.20 &  \\
 & DS & 79.29 & \textbf{0.77} & 0.79 & 0.78 & 0.21 &  \\
 & GR & 80.13 & \textbf{0.77} & 0.80 & 0.78 & 0.20 &  \\
\midrule
\multirow{6}{*}{Attack Vector}
 & G4 & 86.57 & 0.73 & 0.87 & 0.86 & 0.23 & 72.65 \\
 & G5 & 87.96 & 0.81 & 0.88 & \textbf{0.88} & \textbf{0.17} &  \\
 & L  & 79.31 & \textbf{0.82} & 0.80 & 0.79 & 0.67 &  \\
 & GM & \textbf{89.42} & \textbf{0.82} & 0.83 & 0.82 & 0.29 &  \\
 & DS & 79.01 & 0.78 & 0.79 & 0.76 & 0.38 &  \\
 & GR & 83.00 & \textbf{0.82} & 0.83 & 0.82 & 0.29 &  \\
\midrule
\multirow{6}{*}{Privileges Required}
 & G4 & 65.44 & 0.67 & 0.65 & 0.64 & 0.40 & 49.51 \\
 & G5 & \textbf{71.61} & \textbf{0.76} & \textbf{0.72} & \textbf{0.70} & \textbf{0.33} &  \\
 & L  & 58.69 & 0.58 & 0.59 & 0.57 & 0.84 &  \\
 & GM & 69.74 & 0.65 & 0.64 & 0.63 & 0.42 &  \\
 & DS & 58.49 & 0.62 & 0.59 & 0.53 & 0.52 &  \\
 & GR & 64.75 & 0.65 & 0.64 & 0.63 & 0.42 &  \\
\midrule
\multirow{6}{*}{User Interaction}
 & G4 & 84.13 & 0.84 & 0.84 & 0.84 & 0.16 & 63.26 \\
 & G5 & \textbf{88.95} & \textbf{0.89} & \textbf{0.89} & \textbf{0.89} & \textbf{0.11} &  \\
 & L  & 77.67 & 0.78 & 0.78 & 0.77 & 0.23 &  \\
 & GM & 87.85 & 0.82 & 0.82 & 0.82 & 0.18 &  \\
 & DS & 75.18 & 0.76 & 0.75 & 0.73 & 0.25 &  \\
 & GR & 82.58 & 0.82 & 0.82 & 0.82 & 0.18 &  \\
\midrule
\multirow{6}{*}{Scope}
 & G4 & 72.86 & 0.73 & 0.73 & 0.73 & 0.27 & 67.84 \\
 & G5 & \textbf{76.68} & 0.73 & \textbf{0.77} & \textbf{0.77} & \textbf{0.23} &  \\
 & L  & 69.48 & 0.70 & 0.70 & 0.70 & 1.20 &  \\
 & GM & 71.25 & \textbf{0.75} & 0.76 & 0.75 & 0.24 &  \\
 & DS & 70.36 & 0.68 & 0.71 & 0.68 & 0.29 &  \\
 & GR & 76.06 & \textbf{0.75} & 0.76 & 0.75 & 0.24 &  \\
\midrule
\multirow{6}{*}{Confidentiality Impact}
 & G4 & 66.96 & 0.69 & 0.67 & 0.67 & 0.41 & 45.36 \\
 & G5 & \textbf{76.05} & \textbf{0.76} & \textbf{0.76} & \textbf{0.76} & \textbf{0.29} &  \\
 & L  & 56.47 & 0.60 & 0.57 & 0.57 & 0.73 &  \\
 & GM & 74.01 & 0.65 & 0.65 & 0.65 & 0.46 &  \\
 & DS & 60.27 & 0.66 & 0.60 & 0.57 & 0.56 &  \\
 & GR & 65.40 & 0.65 & 0.65 & 0.65 & 0.46 &  \\
\midrule
\multirow{6}{*}{Integrity Impact}
 & G4 & 69.54 & 0.70 & 0.70 & 0.69 & 0.40 & 37.30 \\
 & G5 & \textbf{78.10} & \textbf{0.78} & \textbf{0.78} & \textbf{0.78} & \textbf{0.29} &  \\
 & L  & 55.65 & 0.58 & 0.56 & 0.55 & 0.81 &  \\
 & GM & 75.38 & 0.68 & 0.68 & 0.68 & 0.44 &  \\
 & DS & 56.58 & 0.66 & 0.57 & 0.55 & 0.62 &  \\
 & GR & 68.24 & 0.68 & 0.68 & 0.68 & 0.44 &  \\
\midrule
\multirow{6}{*}{Availability Impact}
 & G4 & 61.84 & 0.57 & 0.62 & 0.57 & 0.52 & 39.65 \\
 & G5 & \textbf{67.95} & \textbf{0.62} & \textbf{0.68} & \textbf{0.62} & \textbf{0.41} &  \\
 & L  & 51.13 & 0.49 & 0.52 & 0.48 & 1.11 &  \\
 & GM & 66.06 & 0.56 & 0.60 & 0.55 & 0.57 &  \\
 & DS & 51.86 & 0.57 & 0.52 & 0.46 & 0.74 &  \\
 & GR & 60.41 & 0.56 & 0.60 & 0.55 & 0.57 &  \\
\midrule
\multirow{6}{*}{Overall}
 & G4 & 73.04 & 0.73 & 0.74 & 0.72 & 0.32 & 57.40 \\
 & G5 & \textbf{78.99} & 0.75 & \textbf{0.79} & \textbf{0.78} & \textbf{0.25} &  \\
 & L  & 65.77 & 0.66 & 0.66 & 0.65 & 0.75 &  \\
 & GM & 76.74 & \textbf{0.77} & 0.77 & 0.76 & 0.28 &  \\
 & DS & 66.38 & 0.67 & 0.67 & 0.63 & 0.45 &  \\
 & GR & 66.38 & 0.71 & 0.72 & 0.71 & 0.35 &  \\
\bottomrule
\end{tabular}
}
\caption{Comparison of model performance across all CVSS v3.1 base metrics. Each block lists six model abbreviations (GPT-4o = G4, GPT-5 = G5, Llama = L, Gemini = GM, DeepSeek = DS, Grok = GR). The Baseline column shows the metric’s baseline accuracy.}
\label{tab:performance_metrics}
\end{table*}

\subsection{Analysis of the content of ``description'' field}
\begin{figure}
    \centering
    \includegraphics[width=\linewidth]{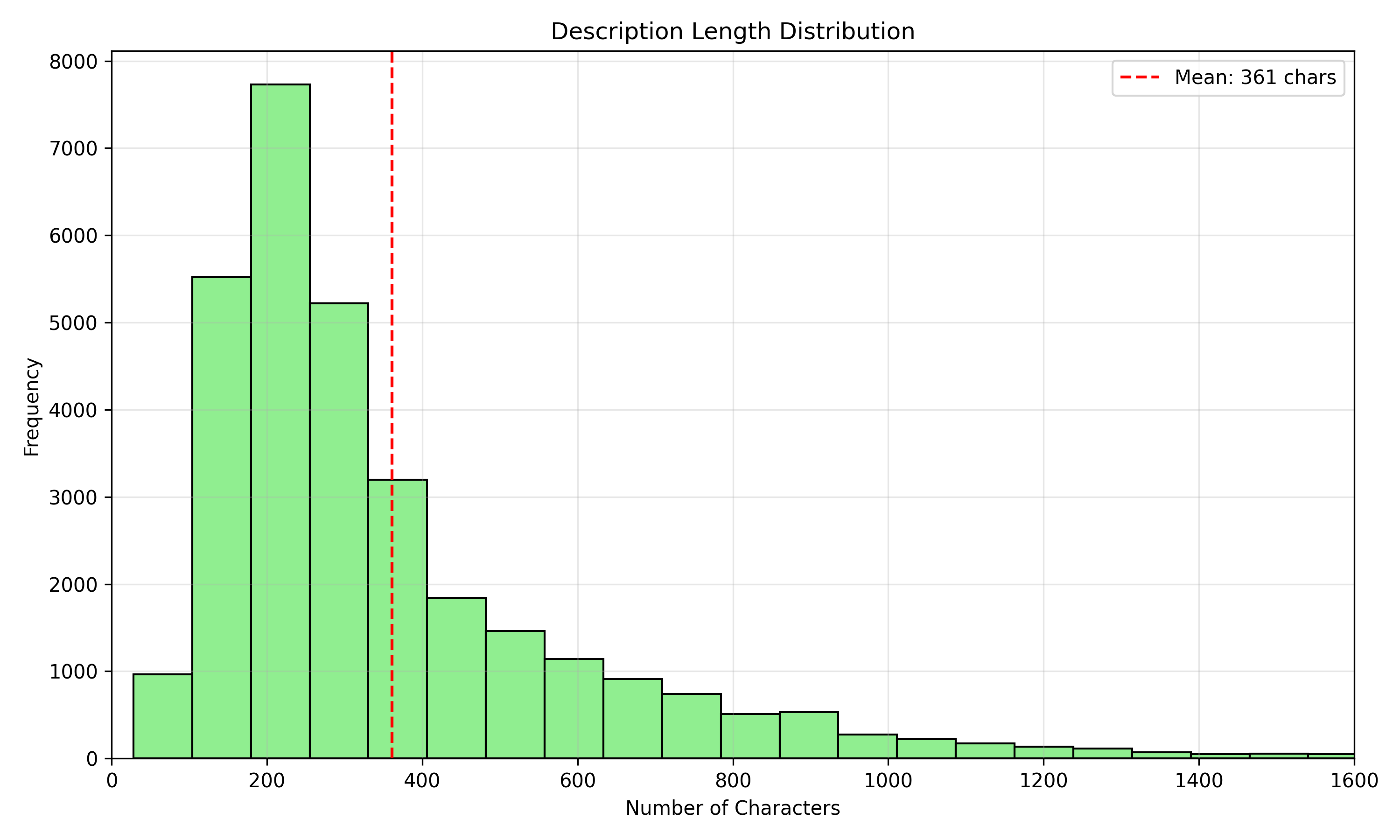}
    \caption{Distribution of Description Length (median=271)} 
    \label{fig:lengh_distribution}
\end{figure}

\noindent\textbf{Length of Description.} The distribution of the length of the descriptions is shown in Figure~\ref{fig:lengh_distribution}. The mean length is 361 characters and the median length is 271. The maximum length of the characters is 3810 and the minimum was 28. The calculated Pearson $r$ indicated no strong correlation between description length and a model's prediction accuracy. 

\noindent\textbf{Named Entities in the Description.} To further assess the interaction of quality of vulnerability descriptions and predictability, we extracted named entities using the \texttt{spaCy} Python library, focusing specifically on entities of type ORG (organizations) and PRODUCT, while excluding PERSON, DATE, and GPE. We then computed the Pearson correlation coefficient between the number of named entities and classification accuracy. The resulting correlation was 0.048 with a $p$-value $<$ 0.05, indicating a weak or negligible relationship between the two variables. There was no  meaningful correlation between the number of named entities in a description and the precision of the classification.

\noindent\textbf{Information Contents.}
We computed the information content (IC) of each description using the Semantic Correlation formula as proposed in prior work on intrinsic IC measures~\cite{seco2004intrinsic}. 
However, we found no strong correlation between IC and the model's precision, suggesting that IC alone is not a determining factor in predictive performance.

\subsection{Meta-LLM Classifiers}
\begin{figure}[H]
    \centering
    \includegraphics[width=\linewidth]{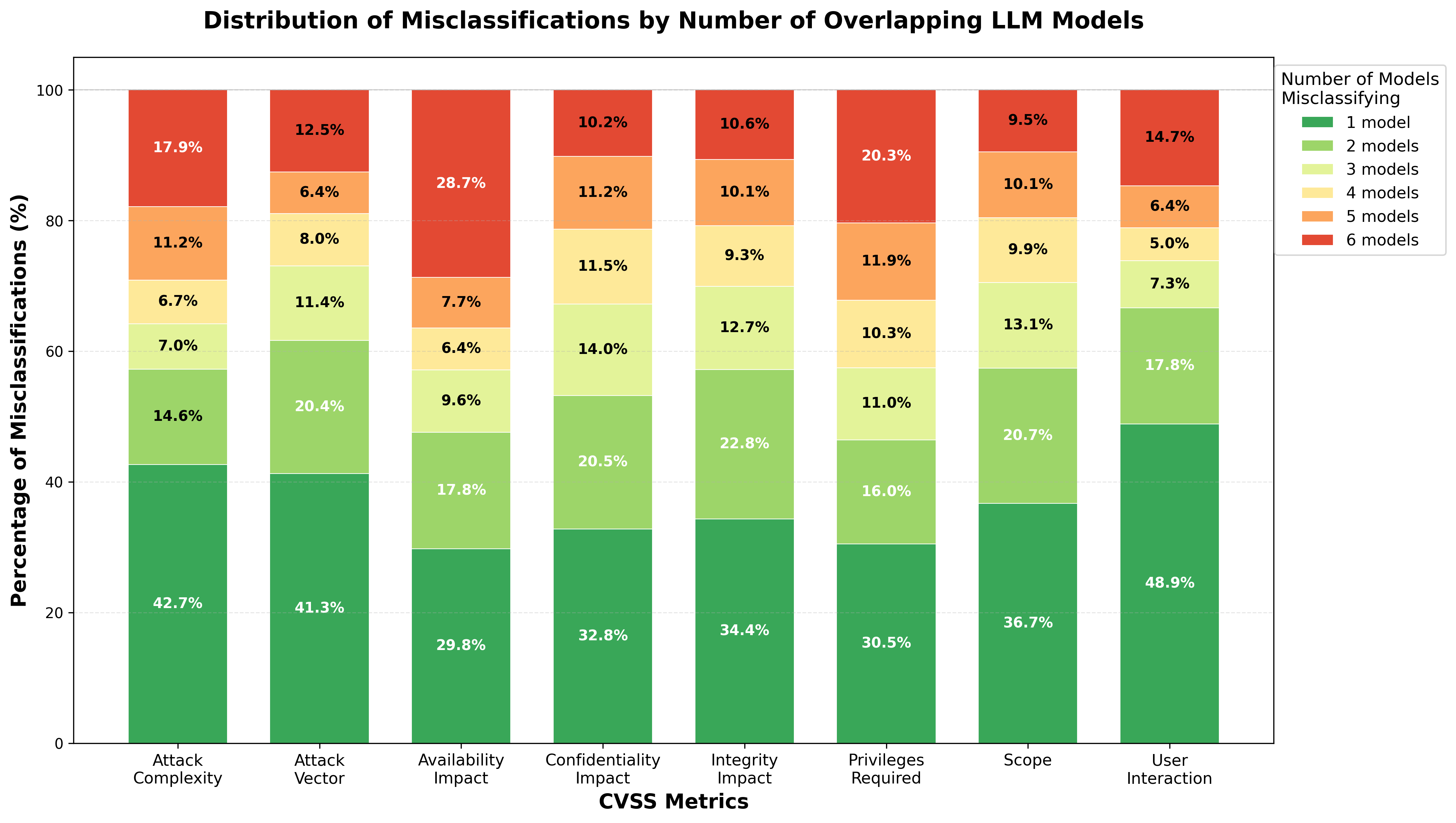}
    \caption{Misclassifications of CVEs across LLMs. Bottom bars show errors by individual models, while red bars indicate CVEs misclassified by all models.} \label{fig:misclassification}
\end{figure}

As shown in Table~\ref{tab:performance_metrics}, different LLMs exhibit varying levels of performance across CVSS metrics. For instance, GPT-5 performs best on \textit{Attack Complexity}, whereas Gemini achieves the highest accuracy on \textit{Attack Vector} classification. To further explore this variability, we analyzed the overlap in misclassifications among the models. Figure~\ref{fig:misclassification} shows that a large portion of misclassifications are shared across models. For instance, all LLMs incorrectly classified 28.7\% of the same CVEs for \textit{Availability Impact}. Likewise, in 17.9\% of the cases, all models misclassified \textit{Attack Complexity}, and in 35.8\% of the cases, the majority of models (four out of six) produced incorrect labels. Nevertheless, we examined whether combining multiple classifiers could enhance performance by exploiting their complementary strengths and mitigating individual errors.

\subsubsection{Feature Engineering}
To enable meta-classification, we constructed a feature engineering pipeline that converts raw LLM predictions into richer representations. The most direct features were the categorical predictions from six LLM models we used, which were encoded numerically to facilitate downstream learning. Beyond these base encodings, we incorporated measures of model agreement, such as the proportion of pairwise consensus among models, the majority-vote class, and the relative strength of that consensus expressed as a confidence score. These agreement-based features allowed the meta-classifier to capture whether a prediction was broadly supported or driven by a single model. Finally, we added simple reliability indicators, including binary flags denoting whether each model produced a valid (non-\texttt{UNKNOWN}) output, with future extensions planned for weighting models based on their historical performance. Together, these features provided a balance of raw predictions, inter-model dynamics, and reliability cues.

\subsubsection{Meta Classifier Architectures}
On top of this feature set, we evaluated several ensemble architectures to determine how best to combine the information for each CVSS metric.The experiments included both traditional and modern classifiers:
\begin{itemize}
    \item \textbf{Voting Ensemble}: A soft-voting method that averaged probability distributions from multiple base learners.
    \item \textbf{Random Forest}: A tree-based approach that provided feature importance analysis.
    \item \textbf{Gradient Boosting}: Another tree-based method, optimized through sequential error correction.
    \item \textbf{Logistic Regression}: A linear model offering interpretable coefficients.
    \item \textbf{Support Vector Machine (SVM)}: Employed with an RBF kernel to capture non-linear decision boundaries in high-dimensional feature space.
    \item \textbf{Neural Network}: A multi-layer perceptron with two hidden layers (100 and 50 neurons), trained with early stopping to prevent overfitting.
\end{itemize} 

This diverse set of models ensured that the meta-classifier evaluation was not tied to any single learning paradigm, but instead explored how different ensemble strategies could exploit the complementary strengths of the underlying LLMs.

\subsubsection{Model Training and Evaluation}
To ensure robust evaluation, we adopted a stratified 5-fold cross-validation strategy that preserved class distributions across folds. Model performance was assessed using standard metrics including accuracy, precision, recall, F1-score, and AUC for binary classification tasks. For each CVSS metric, the model yielding the highest cross-validation F1-score was selected for further testing.

In addition to cross-validation, we employed a conventional train–test split, allocating 80\% of the data for training and reserving 20\% as a held-out test set. Stratified sampling was again applied to maintain representative class distributions across both sets, enabling a fair and consistent evaluation of generalization performance.

\begin{table*}[htbp]
\centering
\caption{Comparison of LLMs vs Best Meta Classifier with Baseline and Change compared to best individual LLM. Meta Models: LR = Logistic Regression, RF = Random Forest, SVM = Support Vector Machine, NN = Neural Network, NM = Naive Bayes Model}
\label{tab:best_comparison}
\begin{tabular}{lcccccccccc}
\toprule
\textbf{CVSS Metric} & \textbf{Baseline} & \textbf{GPT-4} & \textbf{GPT-5} & \textbf{LLaMA} & \textbf{Grok3} & \textbf{Gemini} & \textbf{DeepSeek} & \textbf{Best Meta} & \textbf{Meta Model} & \textbf{Change} \\
\midrule
Attack Complexity & 83.24 & 81.44 & \textbf{83.94} & 76.95 & 79.02 & 79.58 & 78.80 & \textbf{84.19} & LR & +0.24 \\
Attack Vector & 73.16 & 86.81 & 88.29 & 79.10 & 82.70 & \textbf{89.27} & 79.77 & \textbf{90.30} & RF & +1.03 \\
Availability Impact & 39.41 & 60.80 & \textbf{67.08} & 49.64 & 58.82 & 64.97 & 51.54 & \textbf{67.37} & RF & +0.29 \\
Confidentiality Impact & 45.20 & 66.23 & \textbf{75.68} & 55.37 & 64.28 & 73.35 & 59.81 & \textbf{76.45} & SVM & +0.77 \\
Integrity Impact & 37.37 & 68.71 & \textbf{77.71} & 54.45 & 67.07 & 74.73 & 56.34 & \textbf{78.10} & NN & +0.39 \\
Privileges Required & 49.95 & 65.16 & \textbf{71.44} & 58.72 & 63.70 & 69.32 & 58.71 & \textbf{72.23} & SVM & +0.80 \\
Scope & 68.28 & 72.29 & \textbf{76.04} & 69.44 & 75.25 & 71.19 & 70.43 & \textbf{79.12} & NN & +3.08 \\
User Interaction & 64.40 & 83.63 & \textbf{88.36} & 77.19 & 81.85 & 87.02 & 75.55 & \textbf{88.55} & SVM & +0.19 \\
\bottomrule
\end{tabular}
\end{table*}

\subsubsection{Meta-classifier Results}
The performance of the individual LLMs served as a baseline, with GPT-5 generally outperforming other models across most CVSS metrics. However, as shown in Table~\ref{tab:best_comparison}, the meta-classifier consistently exceeded the strongest individual results. For example, Support Vector Machines proved most effective for metrics such as Privileges Required, Confidentiality Impact, and User Interaction, while Random Forest showed gains for Attack Vector and Availability Impact. Neural Networks exceeded model performance for Scope and Integrity Impact, and although the difference was slim, +0.24, Logistic Regression was able to further improve on an already high-scoring metric, Attack Complexity. 

Overall, the meta-classifier achieved an average accuracy of 79.54\%, surpassing GPT-5’s 78.55\%. While the margin of improvement may appear modest, the fact that every CVSS metric benefited from ensemble integration highlights the robustness of the approach. Particularly notable were improvements in Scope (+3.08\%) and Attack Vector (+1.03\%), showing that even high-performing individual metrics could still be enhanced through meta-classification.

These findings suggest that meta-classification provides a systematic advantage by leveraging the complementary strengths of different LLMs while mitigating the risk of poor performance on specific metrics. Although impact-related dimensions (Availability, Confidentiality, and Integrity) remain inherently more challenging, the observed improvements demonstrate that ensemble approaches can progressively enhance the reliability of automated vulnerability assessment. Furthermore, since comparable LLMs tend to make similar errors, providing clearer vulnerability descriptions and richer contextual information can further improve the accuracy and consistency of CVSS scoring automation.

\section{Findings}
\label{s:findings}

Our evaluation of individual LLM models and meta-classifiers for automated CVSS scoring reveals several important findings about their performance, limitations, and the characteristics of the vulnerability descriptions:

\subsection{Performance Across CVSS Metrics}
Among the evaluated models, GPT-5, Gemini, and Grok demonstrated the strongest overall performance across the CVSS base metrics. The highest accuracies were observed for Attack Vector and User Interaction, reflecting strong capabilities in identifying remote access conditions and user involvement cues. In contrast, lower performance on Availability Impact and Privileges Required highlights persistent challenges in impact-related dimensions. The meta-classifier further enhanced overall classification accuracy across all eight metrics by integrating the strengths of all six models, yielding the most significant improvement on the Scope metric.

\subsection{Performance vs. Baseline}
Overall, the classifiers struggled in scenarios with extreme class imbalance—for example, in the case of Attack Complexity, where 83\% of the samples are labeled as \textit{Low}. Even the best-performing LLM, GPT-5, achieved less than a 1\% improvement over the baseline. This highlights the inherent bias that classifiers exhibit when trained on imbalanced data distributions.

\subsection{Quality of Description and Predictive Accuracy}

We used three proxies to assess description informativeness: (1) count of named entities (ORG, PRODUCT), (2) text length, and (3) Information Content. The Pearson correlation between named entities and accuracy for GPT was low ($r = 0.048$, $p < 0.05$). Longer descriptions nor higher IC did not correlate with higher performance, suggesting verbosity is not a strong quality signal.

\subsection{Additional Fields and Predictive Accuracy} 
We investigated whether supplemental contextual features, including CPE, CWE ID, and CVE ID, could improve model performance by comparing them with description-only prompts on a subset of 1,000 CVEs from 2024. Despite providing additional structure, CPE and CWE data produced results comparable to description-only prompts, suggesting that most relevant information is already captured in the textual description. As a result, incorporating additional structured data does not substantially improve classification accuracy. In contrast, adding the CVE ID yielded a large but misleading performance increase, as models used the identifier to retrieve known CVSS vectors rather than infer them. To preserve the integrity of our evaluation, subsequent experiments omitted CWE and CPE fields for redundancy reasons and excluded CVE IDs to prevent lookup bias. The final experimental setup therefore relied solely on human-readable CVE descriptions as model input.

\section{Discussion and Future Work} \label{s:discuss}
Some overlap with pretraining data is possible, since older CVSS records may appear in public corpora. We removed CVE identifiers to limit direct memorization, but partial exposure cannot be ruled out. However, model behavior indicates that recall alone does not drive predictions: performance degrades when key contextual cues are missing. Thus, CVSS scoring depends on contextual reasoning rather than surface-pattern memorization, as models infer well from subtle signals but fail when essential information is absent. To search into this, we asked GPT-4o two questions for a small sample of misclassified CVEs:
\begin{itemize}
    \item Why did you misclassify this CVE description?
    \item What additional information would help predict the correct labels?
\end{itemize}

Main groups of reasons provided by the model for misclassifications include:
\begin{itemize}
    \item \textit{Missing Information:} Descriptions often omit crucial details, such as required privileges, user interaction, or configuration dependencies.
    \item \textit{Ambiguous Language:} Phrases like ``arguments can be provided'' or ``introduce properties'' mislead the model into defaulting to low complexity and no privileges.
    \item \textit{Training Bias:} The model overgeneralizes from similar past examples, applying incorrect heuristics in nuanced cases.
\end{itemize}

Furthermore, despite the improvements in accuracy made by the meta-classifier, the largest change was only +3.08. This suggests that classifiers make no inherent difference to performance, and that enriching CVE descriptions with proper context is the most vital factor for accurate scoring. We plan to augment descriptions with external context ( e.g., libraries involved, dependency graphs, and proof-of-concept exploits where available) to better capture impact and improve model predictions.

\section{Conclusion} 
\label{s:conclude}
This study evaluated the feasibility of using general-purpose LLMs, including GPT-4o, GPT-5, Llama-3.3-70B-Instruct, Gemini-2.5-Flash, DeepSeek-R1, and Grok-3, to automate CVSS v3.1 base metric classification from vulnerability descriptions. Across more than 31,000 CVEs, GPT-5 and Gemini-2.5-Flash showed the strongest overall performance, particularly for metrics such as Attack Vector and User Interaction. However, all models struggled with minority classes and were sensitive to class imbalance and ambiguous descriptions. The meta-classifier offered only modest accuracy improvements, indicating that fine-tuning and domain-specific enhancements remain necessary. Overall, our findings show that LLMs can support scalable vulnerability triage but also reveal key limitations that must be addressed. Future work should explore instruction tuning and the integration of external contextual signals to further improve prediction reliability in operational settings. All project code and data have been released as open-source materials~\cite{ecdeans_AI-Vuln-Management}.


\sloppy {
\printbibliography
}



\end{document}